\documentclass[preprint,sort&compress]{elsarticle}
\usepackage{graphicx}
\usepackage{float}
\usepackage[dvips]{color}
\def\bc{\begin{center}}
\def\ec{\end{center}}
\def\vs{\vspace{-0.5cm}}
\def\beq{\begin{equation}}
\def\eeq{\end{equation}}

\begin{document}

\author[rvt]{O. Fialko }
\ead{Oleksandr.Fialko@physik.uni-augsburg.de}
\author[els]{A. Gammal \fnref{fn2}}
\ead{gammal@if.usp.br}
\author[rvt]{K. Ziegler }
\ead{klaus.ziegler@physik.uni-augsburg.de}

\fntext[fn2]{Phone: +55-11-3091-6659, FAX: +55-11-3091-6832}

\address[rvt]{Institut f\"ur Physik, Universit\"at Augsburg, D-86135, Germany}
\address[els]{Instituto de F\'{\i}sica, Universidade de S\~ao Paulo, 05508-090, S\~ao Paulo-SP, Brazil}


\title{A Strongly Attractive Fermi Gas in an Optical Lattice}

\begin{abstract} 
We study strongly attractive fermions in an optical lattice superimposed by a trapping potential.
We calculate the densities of fermions and condensed bound molecules at
zero temperature. There is a competition between dissociated fermions and molecules 
leading to a reduction of the density of fermions at the trap center.
\end{abstract}


\maketitle

\section{Introduction}
Ultra-cold gases present many-body systems with a remarkable tunability of its parameters,
providing us with a
platform for the investigation of numerous properties of complex quantum systems. For instance,
it has become possible to tune the interaction strength between atoms over a wide range by
Feshbach resonance \cite{timmermans99}.
The investigation of ultra-cold Fermi gases started shortly after the discovery of Bose-Einstein condensation
of bosonic atoms \cite{anderson95} when quantum degeneracy in a gas of fermionic atoms was obtained
\cite{jin99}. For fermions with weak attraction we have the celebrated phenomenon of Cooper pairing
\cite{cooper57}. By tuning to strong
attraction one enters the regime of diatomic molecules, whose size is much smaller than that of Cooper pairs. These
molecules then may condense into a Bose-Einstein condensate of a hard-core Bose gas.
The crossover from the weakly interacting BCS regime to the strongly interacting BEC regime
has been the subject of theoretical studies in \cite{leggett80,nozieres85}.
The bosonic molecules, formed of pairs of fermionic atoms, were produced experimentally in a trapped system 
\cite{reagal03} as well as in an optical lattice \cite{stoferle06}. 

Most investigations were based on continuous Fermi gases. The more recently introduced optical lattices in 
ultra-cold
gases \cite{greiner01} may have a number of interesting effects on the Fermi gas. First of all, the
dispersion of the atoms will be changed by the lattice. Moreover, the interaction between the atoms has a strong
effect on the quantum
states of the Fermi gas by allowing, for instance, to form Mott states \cite{fisher89}. This could mean
that the molecular gas in the BEC regime does not condense but becomes a Mott state. In the BCS regime, on the other
hand, the effect of an optical  lattice is not so dramatic because the Cooper pair radius is much larger than the
lattice spacing.  Consequently, the BEC-BCS crossover can be much richer in the presence of an optical lattice. 

In this paper, we study strongly attractive fermions in an optical lattice superimposed by a trapping potential. 
We first show that the phase diagram of tightly bound fermions contains a Bose-Einstein condensed phase
and a Mott insulating phase of such molecules. Then we study the system in a harmonic trap and calculate the density 
of unpaired fermions in the presence of a condensate state of molecules. We show that there is a competition between paired 
fermions and unpaired fermions which leads to a reduction of the density of unpaired fermions at the center of the 
trap.

\section{Model}

The simplest lattice model of an interacting Fermi gas is the Hubbard model \cite{lewenstein07}.
It describes the competition between the kinetic energy of the fermions and a local interaction
and provides a phase diagram that includes a Fermi liquid and a Mott insulator. Originally introduced
for a repulsive Fermi gas (e.g. electrons in a metal), the model can also be used for neutral fermionic
atoms with attractive local interaction. For very strong attraction, however, the local interaction
is insufficient to describe the physics of the Fermi gas. This due to the fact that strong attraction
causes pairing of fermions to local molecules. On the other hand, the kinetic term in the Hubbard model
allows only individual tunneling of fermions. This means that the tightly bound molecules must dissociate
into independent fermions in order to tunnel in the optical lattice. The associated energy of such
a process is of the order of the attractive interaction. (Actually, the effective tunneling rate
is $\sim 2t^2/U$ \cite{nozieres85}, where $t$ is the tunneling rate of individual fermions and $U$ is
the strength of the local attraction.) This means that the motion of molecules is strongly suppressed
in the Hubbard model \cite{orso05}. On the other hand, there is no reason for the molecules not tunnel freely in
the optical lattice because they only have to obey the Pauli principle. The solution of this problem
is an extension of the attractive fermionic Hubbard model that includes an additional kinetic term for
the bosonic molecules \cite{duan05,carr05,ziegler05,ohashi08}. The Hamiltonian for an attractive Fermi gas
in a $d$-dimensional optical lattice is a molecular fermionic Hubbard (MFH) model and reads
\beq
\hat{H}_f=-\frac{t}{2d}\sum_{\sigma=\uparrow,\downarrow}\sum_{\langle
r,r'\rangle}\hat{c}_{r\sigma}^{\dagger}\hat{c}_{r'\sigma}
-\frac{J}{2d}\sum_{\langle
r,r'\rangle}\hat{c}_{r\uparrow}^{\dagger}\hat{c}_{r'\uparrow}\hat{c}_{r
\downarrow}
^{\dagger}\hat{c}_{r'\downarrow}
-U\sum_{r}\hat{c}_{r\uparrow}^{\dagger}\hat{c}_{r\uparrow}\hat{c}_{r\downarrow}
^{\dagger}\hat{c}_{r\downarrow}
- \sum_{\sigma=\uparrow,\downarrow}\sum_{r}\mu_\sigma
\hat{c}_{r\sigma}^{\dagger}\hat{c}_{r\sigma}.
\label{model}
\end{equation}
Here, $\hat{c}_{r,\sigma}$ ($\hat{c}_{r,\sigma}^\dagger$)
is the annihilation (creation) operator for particles at lattice
site $r$. The index $\sigma=\uparrow,\downarrow$ represents two hyperfine states of fermionic atoms, e.g., $^{40}$K or
$^{6}$Li. In this work we consider only the symmetric case where the number of fermions in each component 
is the same.
Nearest-neighbor tunneling of the individual fermions is described by the parameter $t$.
There is also a term with parameter $J$ which is understood as a tunneling term of dressed fermionic pairs 
\cite{duan05, ohashi08}. 
$U\sim U_{bg} - 4dg^2/\delta$ accounts for an effective local attractive interaction between fermions with the 
detuning $\delta$,  coupling between fermions and molecules $g$ 
and $U_{bg}=4\pi\hbar^2a_b/m$ with the background scattering length 
$a_b$ and $m$ is the mass of fermions. In the strong coupling regime $|\delta|\gg g$ and $a_b$ is small and positive 
\cite{duan05}. So that for strong attractions $U < J$.
A parabolic trapping potential $V_r=\gamma(x^2+y^2)$ (we will study a 2D system) can be combined with the 
chemical potential 
$\mu_{\sigma}$ to $\mu_{r\sigma}=\mu_{\sigma}-V_r$, which controls the number of particles in a grand-canonical 
ensemble, $\gamma$ is the strength of the trap.

\section{Functional integral representation}

The grand-canonical ensemble is given by the partition function as a functional integral with respect to the
Grassmann fields, ${\cal{Z}}=\int e^{-S(\bar{\psi},\psi)}D[\bar{\psi},\psi]$,
with the action \cite{simons06}
\beq
S(\bar{\psi},\psi)=\int_0^{\beta}d\tau\left[\sum_r
(\bar{\psi}_{r\tau\uparrow}\partial_{\tau}\psi_{r\tau\uparrow}+\bar{\psi}_{r\tau\downarrow}\partial_{\tau}\psi_
{r\tau\downarrow})
+H_f(\bar{\psi}_{\tau+\epsilon},\psi_{\tau})\right],
\eeq
where $\epsilon$ should be sent to $+0$ at the end of calculations, $\tau$ is the imaginary time, the symbol
$\partial_{\tau}\psi_{\tau}$ denotes the formal $\lim_{\epsilon\rightarrow +0}(\psi_{\tau+\epsilon}-\psi_{\tau})$,
$\beta\equiv (k_B T)^{-1}$ is the inverse temperature. $H_f(\bar{\psi}_{\tau+\epsilon},\psi_{\tau})$ is obtained
by replacing creation and annihilation operators in the Hamiltonian $\hat{H}_f$ by Grassmann fields $\bar{\psi}$
and $\psi$ respectively.
We will perform a Hubbard-Stratonovich transformation to decouple the fourth order terms at the expense
of introducing new complex fields. 
We use the identity
\[
\exp\left(\frac{J}{2d}\sum_{\langle
r,r^{\prime}\rangle}\bar{\psi}_{r\tau+\epsilon\uparrow}\psi_{r^{\prime}\tau\uparrow}
\bar{\psi}_{r\tau+\epsilon\downarrow}\psi_{r^{\prime}\tau\downarrow}
+U\sum_{r}\bar{\psi}_{r\tau+\epsilon\uparrow}\psi_{r\tau\uparrow}
\bar{\psi}_{r\tau+\epsilon\downarrow}\psi_{r\tau\downarrow}\right)
\]
\vs
\[
\propto\int[d\phi][d\chi]\exp\left(
-\sum_{r,r^{\prime}}\bar{\phi}_{r\tau}\hat{v}^{
-1}_{r,r^{\prime}}\phi_{r^{\prime}\tau}-IU^{-1}\sum_{r}\bar{\chi}_{r\tau}\chi_{r\tau}
\right.
\]
\beq
\left. 
-\sum_r (i\phi_{r\tau}+I\chi_{r\tau})\bar{\psi}_{r\tau+\epsilon\uparrow}\bar{\psi}_{r\tau+\epsilon\downarrow}
-\sum_r
(i\bar{\phi}_{r\tau}+I\bar{\chi}_{r\tau})\psi_{r\tau\uparrow}\psi_{r\tau\downarrow}\right),
\eeq
where $ I=\theta(J-|U|),
\hat{v}_{r,r^{\prime}}=\frac{J}{2d}\delta_{|r-r^{\prime}|,a}-J\delta_{r,r^{\prime}}
+(1+I)U\delta_{r,r^{\prime}}$.
The subsequent integration over Grassmann fields leads to the effective action in terms of the
two complex fields $\phi$, $\chi$ and their conjugate fields $\bar{\phi}$, $\bar{\chi}$
\beq
\label{action1}
S_{eff}=
\int_0^{\beta}d\tau\left[
\sum_{r,r^{\prime}}\bar{\phi}_{r\tau}\hat{v}^{-1}_{r,r^{\prime}}\phi_{r^{\prime}\tau}
+IU^{-1}\sum_r \bar{\chi}_{r\tau}\chi_{r\tau}
-\ln\det \hat{G}^{-1} \right],
\eeq
with an inverse Nambu-Gor'kov propagator
\begin{equation}
\hat{G}^{-1}=\left(\begin{array}{cc}-i\phi-I\chi & \partial_{\tau}+\mu_{\downarrow}+\hat{t} \\
\partial_{\tau}-\mu_{\uparrow}-\hat{t} &
i\bar{\phi}+I\bar{\chi} \end{array}\right).
\label{Green}
\end{equation}
Here $i\phi+I\chi$ is a diagonal matrix with $i\phi_{r\tau}+I\chi_{r\tau}$ on its diagonal.
$\mu_{\sigma}$ is also a diagonal matrix with elements
$\mu_{r\sigma}$. $\hat{t}$ is a hopping matrix with elements $t/2d$.

The total density of fermions can be calculated as a derivative of the free energy 
$F=\ln {\cal{Z}}/\beta$ with respect to the chemical potential, $n_{r\sigma}=\partial F/\partial \mu_{r\sigma}$.
In the strongly interacting regime in an optical lattice we expect that the size of a
molecule is within one well of the optical lattice and the internal structure is irrelevant.
The corresponding creation operator is  $c^{\dagger}_{r\uparrow}c^{\dagger}_{r\downarrow}$.
Then the molecules can be treated as point-like objects.  The scattering length of fermions is
assumed to be
less than the lattice spacing of the optical lattice, since the size of the molecules is of the order of
the scattering length.

In order to calculate the condensate density we consider the presence
of ``off-diagonal long-range order" \cite{yang62}
\beq
n_0=\lim_{|r-r^{\prime}|\rightarrow \infty}
\frac{1}{\beta}\int d\tau \lim_{\epsilon\rightarrow +0} \langle \bar{\psi}_{
r\tau+\epsilon\uparrow}\psi_{ r^{\prime}\tau\uparrow}\bar{\psi}_{
r\tau+\epsilon\downarrow}\psi_{r^{\prime}\tau\downarrow} \rangle.
\eeq
It can be shown that the latter is related to the correlations of the complex fields,
$n_0\sim \lim_{|r-r^{\prime}|\rightarrow \infty}\langle \phi_r\bar{\phi}_{r^{\prime}} \rangle
\label{assimp}$, where $\langle\dots\rangle={\cal{Z}}^{-1}\int\ldots e^{-S_{eff}}D[\phi,\chi]$
\cite{moseley08}.

\begin{figure}[t]
\begin{center}
\includegraphics[width=10cm]{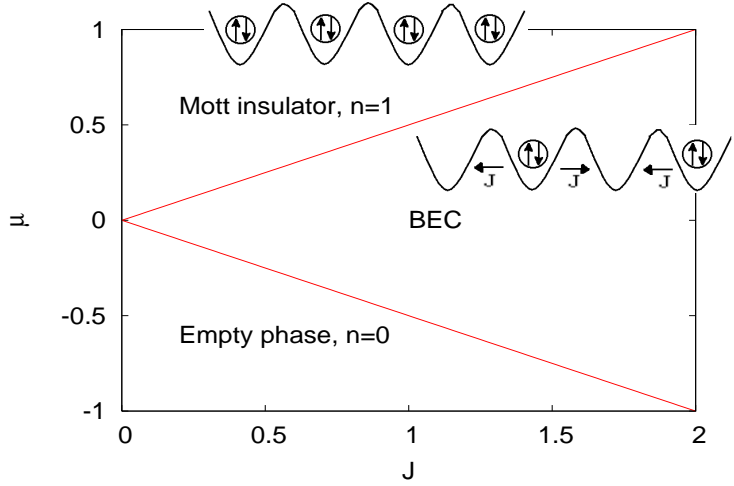}
\end{center}
\caption{Phase diagram for $t=0$, $U=0$ and $k_B T=0$.
The three phases are the BEC with a non-vanishing condensed density, the MI
states with vanishing condensed density and with one molecules per lattice site and the empty phase.
$\mu$ and $J$ are in arbitrary energy units.}
\label{diagram}
\end{figure}
In the limiting case of strong attraction all fermions are paired and the tunneling of molecules occurs
without their dissociation. In this case $U$ can be neglected and the atomic tunneling rate $t$ is not effective. 
Without a harmonic trap ($\gamma=0$), the integration over the complex molecular fields $\phi,\chi$ can be 
performed in saddle-point approximation \cite{fialko07}. 
The phase diagram is depicted in Fig. \ref{diagram}. There are three phases: the BEC
of molecules with the condensed fraction $n_0=(J^2-4\mu^2)/(4J^2)$, the MI state with one 
particle per site and $n_0=0$ and the empty phase (cf. \cite{carr05}).
Fluctuations around the saddle point provide the low-energy excitations of the bosonic molecules. They are gapless
in the BEC phase
\beq
\epsilon_{{k}}=\sqrt{4J^2g_{{k}}n_0 +4g_{{k}}^2\mu^2},
\label{spectrum0}
\eeq
where $g_{{k}}=1-1/d \sum_{i=1}^d \cos (k_ia)$ is the dispersion of the free Bose gas.
On the other hand, the excitations of the MI state have a gap $\tilde{\Delta}=2\mu-J>0$ 
\beq
\epsilon_{{k}}=\tilde{\Delta}+Jg_{{k}}
\ .
\eeq

\section{Ground state of the trapped system}

The approximate ground state of the system is obtained as the saddle point of the action in Eq. 
(\ref{action1}) with respect to the fields (i.e., by solving the equation $\delta S_{eff}=0$). 
Here we assume that the solution is static (i.e. independent of $\tau$). Moreover, we also use 
the fact that a slowly varying field $\phi$ in space of a trapped condensate can be 
approximated as \begin{equation} 
\sum_{r^{\prime}}\hat{v}_{r,r^{\prime}}^{-1}\phi_{r^{\prime}}\approx 
b\phi_r+Jb^2\sum_{r^{\prime}}\left(\delta_{rr^{\prime}}-\hat{J}_ {rr^{\prime}} 
\right)\phi_{r^{\prime}}, \label{approxident} 
\end{equation} 
where $b^{-1}=(1+I)U$, 
$\hat{J}$ is a tunneling matrix with elements $1/2d$. Then we obtain from $\delta S_{eff}=0$ 
the following equations for the fields
$\phi$, $\bar{\phi}$ 
\beq 
\label{E1} -{Jb^2\over 
2d}\sum_{l=1}^d({\bar\phi}_{r+ae_l}-2{\bar\phi}_{r} +{\bar\phi}_{r-ae_l}) = 
-b{\bar\phi}_r-{\rm i}\frac{1}{\beta}\sum_n G_{rr,11}(\omega_n), 
\eeq
\beq \label{E2} -{Jb^2\over 2d}\sum_{l=1}^d(\phi_{r+ae_l}-2\phi_{r}+\phi_{r-ae_l}) = 
-b\phi_r+{\rm i}\frac{1}{\beta}\sum_nG_{rr,22}(\omega_n) \ , \eeq
where $e_l$ is the Cartesian lattice unit vector in direction $l$. $G_{rr',11-12-21-22}$ are 
defined as the $2\times2$ block structure of
\begin{equation} G=\left(\begin{array}{cc} G_{rr',11} & G_{rr',12} \\ G_{rr',21} & G_{rr',22} 
\\ \end{array}\right), \label{Gr1112} \end{equation}
$r=1,2,...,N^d$, where $G$ is the inverse matrix of \begin{equation} 
G^{-1}=\left(\begin{array}{cc} -i\phi-I\chi & i\omega_n+\mu_{\downarrow}+\hat{t} \\ 
i\omega_n-\mu_{\uparrow}-\hat{t} & i\bar{\phi}+I\bar{\chi} \end{array}\right). 
\label{GreenFourier}
\end{equation} 
$\omega_n=\pi(2n+1)/\beta$ is a Matsubara frequency originating from the Fourier 
transformation $\partial_{\tau}\rightarrow -i\omega_n$. The equations of the other complex 
field $\chi$ read 
\begin{equation}
{\bar\chi}_r=-IU\frac{1}{\beta}\sum_n G_{rr,11}(\omega_n), 
\label{E3} 
\end{equation} 
\begin{equation} 
\chi_r=IU\frac{1}{\beta}\sum_n G_{rr,22}(\omega_n) \ . 
\label{E4} 
\end{equation}
Eqs. (\ref{E1}) and (\ref{E2}) are analogous to the Gross-Pitaevskii equation for a Bose gas 
and provide a macroscopic wave function of the condensate molecules. The left-hand side of Eqs. 
(\ref{E1}) and (\ref{E2}) can be understood as a lattice Laplacian acting on the field
\beq \nabla^2 {\bar\phi}_r= \sum_{l=1}^d({\bar\phi}_{r+ae_l}-2{\bar\phi}_{r}+{\bar\phi}_{r-ae_l}) \ . \label{laplacian} 
\eeq
It should also be noticed that the additional condition $\nabla^2 \phi=0$ gives the Thomas-Fermi 
approximation. 
Using the effective action in Eq. (\ref{action1}) and differentiating it
with respect to the chemical potential we get the total densities
\beq
n_{r\uparrow}=\frac{1}{\beta}\sum_n G_{rr,12}(\omega_n),
\hskip0.5cm
n_{r\downarrow}=-\frac{1}{\beta}\sum_n G_{rr,21}(\omega_n) \ .
\label{total}
\eeq
In mean-field approximation (cf. Ref. \cite{fialko07}) this gives for the densities of unpaired fermions
\beq
n^f_{r\uparrow}=n_{r\uparrow}-n_{r\uparrow}n_{r\downarrow}-n_{0,r}, \ \
n^f_{r\downarrow}=n_{r\downarrow}-n_{r\uparrow}n_{r\downarrow}-n_{0,r}.
\label{disatoms}
\eeq
Here $n_f$ measures the presence of a fermion and absence of a pair
of fermions. $n_{r\uparrow}n_{r\downarrow}$ gives the probability to find two
fermions at a lattice site $r$. Finally, $n_{0,r}$ is the product of anomalous averages $\sim \langle\psi_{\uparrow}
\psi_{\downarrow}\rangle\langle\bar{\psi}_{\downarrow}
\bar{\psi}_{\uparrow}\rangle$ and thus can be associated with the condensed density
\beq
n_{0,r}=\frac{1}{\beta}\sum_n G_{rr,11}(\omega_n)
\frac{1}{\beta}\sum_n G_{rr,22}(\omega_n)\sim |i\phi_r+\chi_r|^2.
\eeq

\section{Numerical calculation}

We calculate numerically the densities of Fermi gas in a 2D dimensional $N\times N$ ($N=30$) sites in an 
optical lattice.
We explored the case $J > |U|$ that corresponds to the strongly interacting regime \cite{duan05}.
We choose the lattice constant $a=0.3$ in our calculation. 
Equations (\ref{E1}) and (\ref{E2}) are elliptical equations and similar to the
Poisson equation. A well known 
technique to solve the Poisson equation numerically is the relaxation method, where an 
artificial time dependent term is added \cite{press}. In two dimensions, for equations 
(\ref{E1},\ref{E2},\ref{E3},\ref{E4}) we obtain the following relaxation equations
\begin{equation} \frac{\partial \bar{\phi}_{i,j}}{\partial t} 
=\bar{\phi}_{i+1,j}-2\bar{\phi}_{i,j}+\bar{\phi}_{i-1,j} 
+\bar{\phi}_{i,j+1}-2\bar{\phi}_{i,j}+\bar{\phi}_{i,j-1} -\frac 
{2d}{Jb}\bar{\phi}_{i,j}-{\rm i}\frac{2d}{Jb^2} \frac{\bar{\chi}_{i,j}}{IU}, \label{R1} 
\end{equation}
\begin{equation} \frac{\partial \phi_{i,j}}{\partial t} 
=\phi_{i+1,j}-2\phi_{i,j}+\phi_{i-1,j} 
+\phi_{i,j+1}-2\phi_{i,j}+\phi_{i,j-1} -\frac {2d}{Jb}\phi_{i,j}+{\rm 
i}\frac{2d}{Jb^2} \frac{\chi_{i,j}}{IU}. \label{R2} 
\end{equation}
\beq 
\frac{\partial{\bar\chi}_r}{\partial t}= -{\bar\chi}_r-IU\frac{1}{\beta}\sum_n 
G_{rr,11}(\omega_n), \label{R3} 
\eeq 
\beq \frac{\partial\chi_r}{\partial 
t}=-\chi_r+IU\frac{1}{\beta}\sum_n G_{rr,22}(\omega_n) \ . \label{R4} 
\eeq
where $i,j=1,2,...,N$ is a Cartesian mapping to $r=1,2,...,N^2$ that describes the coordinates 
of the square lattice.
Eqs. (\ref{R1}) and (\ref{R2}) are evolved through a splitting operator technique \cite{press}. 
The corresponding $\nabla^2$ terms are evolved by a Crank-Nicolson algorithm while the remaining 
ordinary equations are evolved by Euler method. 
The equations are evolved until 
$\bar{\phi},\phi,\bar{\chi},\chi$ no longer changes with time, i.e., $\partial 
\bar{\phi}/\partial t=\partial \phi/\partial t= \partial \bar{\chi}/\partial t =\partial 
\chi/\partial t=0$ and thus the set of equations are equation is satisfied.
Equations (\ref{R1},\ref{R2},\ref{R3},\ref{R4}) where discretized with time steps ranging from 
0.01 to 0.001. 
We start the evolution of eqs. (\ref{R1}) 
and (\ref{R2}) from an initial Gaussian ansatz for $\bar{\phi}$, $\phi$ and 
$\bar{\chi}=\chi={\rm i}$. We then apply a numerical diagonalization to equations described in 
the appendix \ref{summat} and obtain the summations $1/\beta\sum_nG_{rr,11}(\omega_n)$ and 
$1/\beta\sum_nG_{rr,12}(\omega_n)$. Then we evolve $\bar{\phi},\phi$ with the $\nabla^2$ terms of 
eqs. (\ref{R1},\ref{R2}), evolve $\bar{\chi},\chi$ in eq. (\ref{R3},\ref{R4}) and finally 
evolve $\bar{\phi},\phi$ with the remaining terms of eq. (\ref{R1},\ref{R2}). These updated 
values will then be used as new ansatz for numerical diagonalization and the circle is repeated. 
Convergence was achieved after $t\sim 10$, or 1000 steps.
The above numerical procedure gives the complex fields $\phi$ and $\chi$ as a function
at each site $r$, i.e., the field at each point in the Cartesian grid $i,j$ in the plane $x,y$, 
where $x=(i-N/2-1/2)a$, $y=(j-N/2-1/2)a$, 
$i,j=1,2,...N$.

\begin{figure}[t]
\begin{center}
\includegraphics[width=10cm]{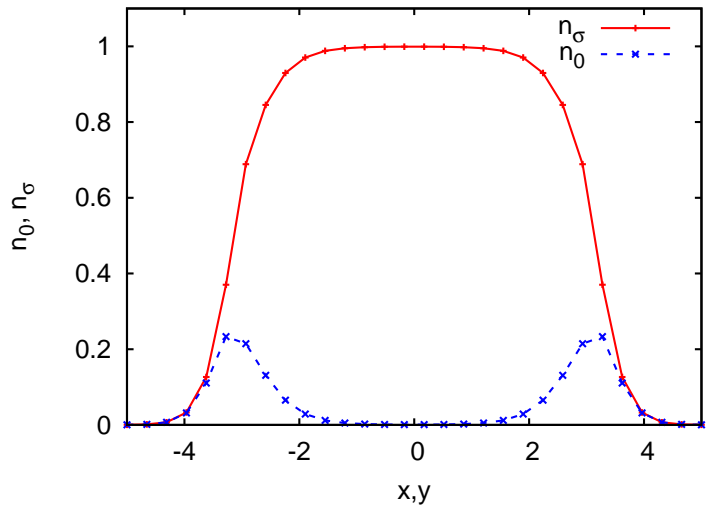}
\end{center}
\caption{Mott state in the trap for values of $t/J=0$ and $U=0$. The chemical potential
is $\mu_{\sigma}=J$. The superfluid shell assembles around the Mott
plateau. 
The density profiles are along the lines $y=a/2$ and $x=a/2$.} 
\label{MOTT}
\end{figure}

First we consider a situation when the Hamiltonian contains only the tunneling of paired fermions, i.e., for  
$t=0$ and $U=0$. This corresponds to 
the limiting case of 
the strongly interacting regime \cite{duan05, ohashi08}. We also fix
$\gamma=0.05 J$ and large chemical potential $\mu_{\sigma}=J$ (large number of fermions).
In Fig. \ref{MOTT} a Mott plateau of the paired molecules is shown. 
The condensed molecules ($n_0$) form a superfluid shell, which assembles around the Mott plateau (cf. 
\cite{kawakami09}).

\begin{figure}[t]
\begin{center}
\includegraphics[width=10cm]{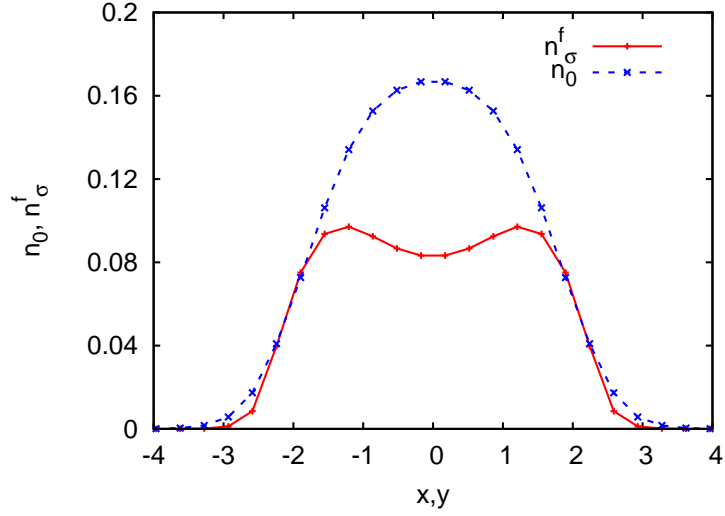}
\end{center}
\caption{The condensate density and the density of dissociated atoms  for $t/J=0.3$,
$U=0.5J$. There is a dip at the center of the trap in the case of the
dissociated atoms. It maybe explained by the Pauli exclusion principle acting between paired 
and unpaired fermions.
The density profiles are along the lines $y=a/2$ and $x=a/2.$} 
\label{dens-U}
\end{figure}

\begin{figure}[h!!]
\begin{center}
\includegraphics[width=10cm]{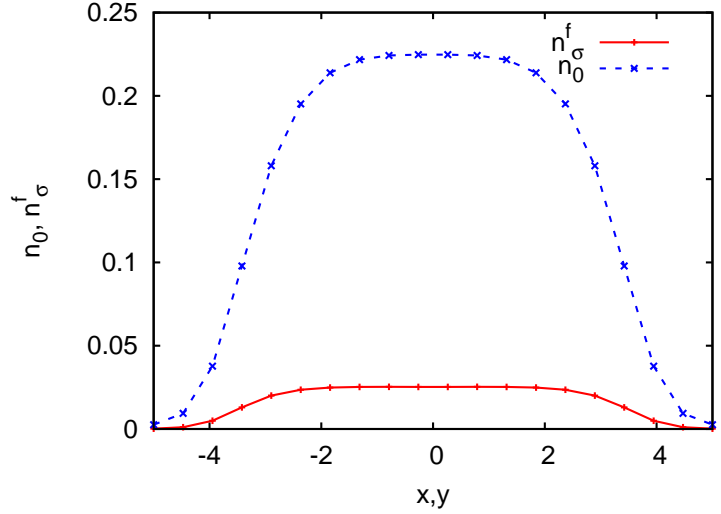}
\end{center}
\caption{The condensate density and the density of dissociated atoms for $t/J=0.5$,
$U=1.5J$. There is no dip at the center of the trap in the case of the
dissociated atoms. The density profiles are along the lines $y=a/2$ and $x=a/2$.}
\label{densU}
\end{figure}

\begin{figure}[t]
\begin{center}
\includegraphics[width=10cm]{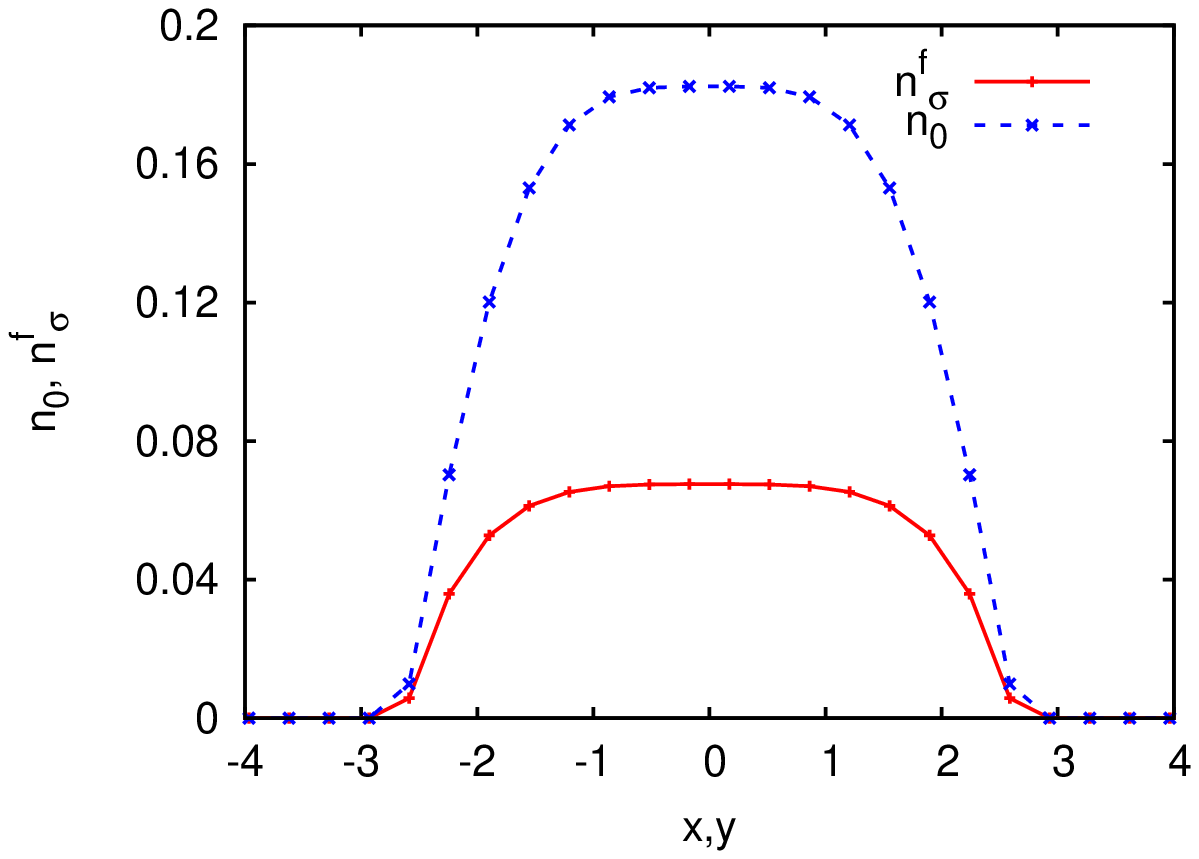}
\end{center}
\caption{The condensate density and the density of dissociated atoms in Thomas-Fermi approximation
for the same values of $t/J$ and $U$ as in Fig. \ref{dens-U}. The Thomas-Fermi approximation
can not reproduce the dip formation. The density profiles are along the lines $y=a/2$ and $x=a/2$.}
\label{TF}
\end{figure}

Now we fix the smaller chemical potential $\mu_{\sigma}=0$ and nonvanishing $t$ and $U$.
In Fig. \ref{dens-U} the condensed density as well as the density of dissociated atoms are plotted for $U=0.5J$. 
We notice that there is a dip at the center of the trap for the
dissociated atoms.
If the condition for the strongly interacting regime is not satisfied
(e.g., $U=1.5J$ in  Fig. \ref{densU}), then this dip is less
pronounced. In Fig. \ref{TF} we also plot the densities in the Thomas-Fermi approximation (by putting terms on the 
left hand sides in Eqs. (\ref{E1}) and (\ref{E2}) to zero)  that does 
not reveal the dip formations as compared to Fig. \ref{dens-U}.

\begin{figure}[h!!]
\begin{center}
\includegraphics[width=10cm]{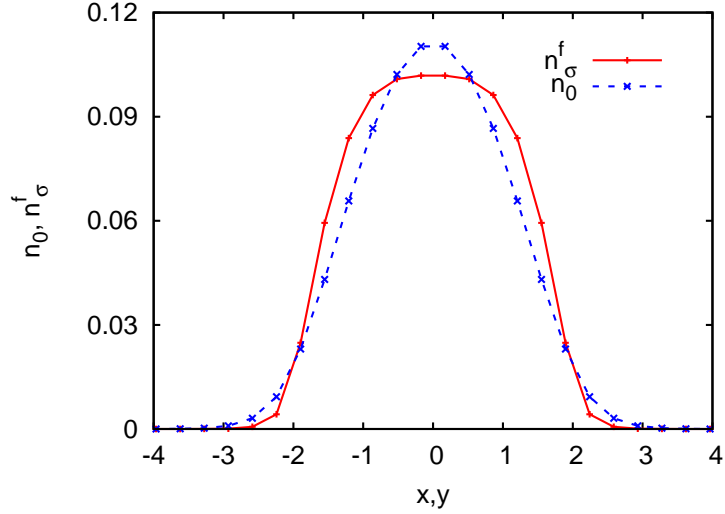}
\end{center}
\caption{The condensate density and the density of dissociated atoms
for the same values of $t/J$ and $U$ as in Fig. \ref{dens-U} but for lower densities with
$\mu_{\sigma}=-0.1J$.
The dip disappears since at lower densities the role of the lattice is less profound as for higher densities.
The density profiles are along the lines $y=a/2$ and $x=a/2$.}
\label{fig:mu-0.1}
\end{figure}

\begin{figure}[t]
\begin{center}
\includegraphics[width=10cm]{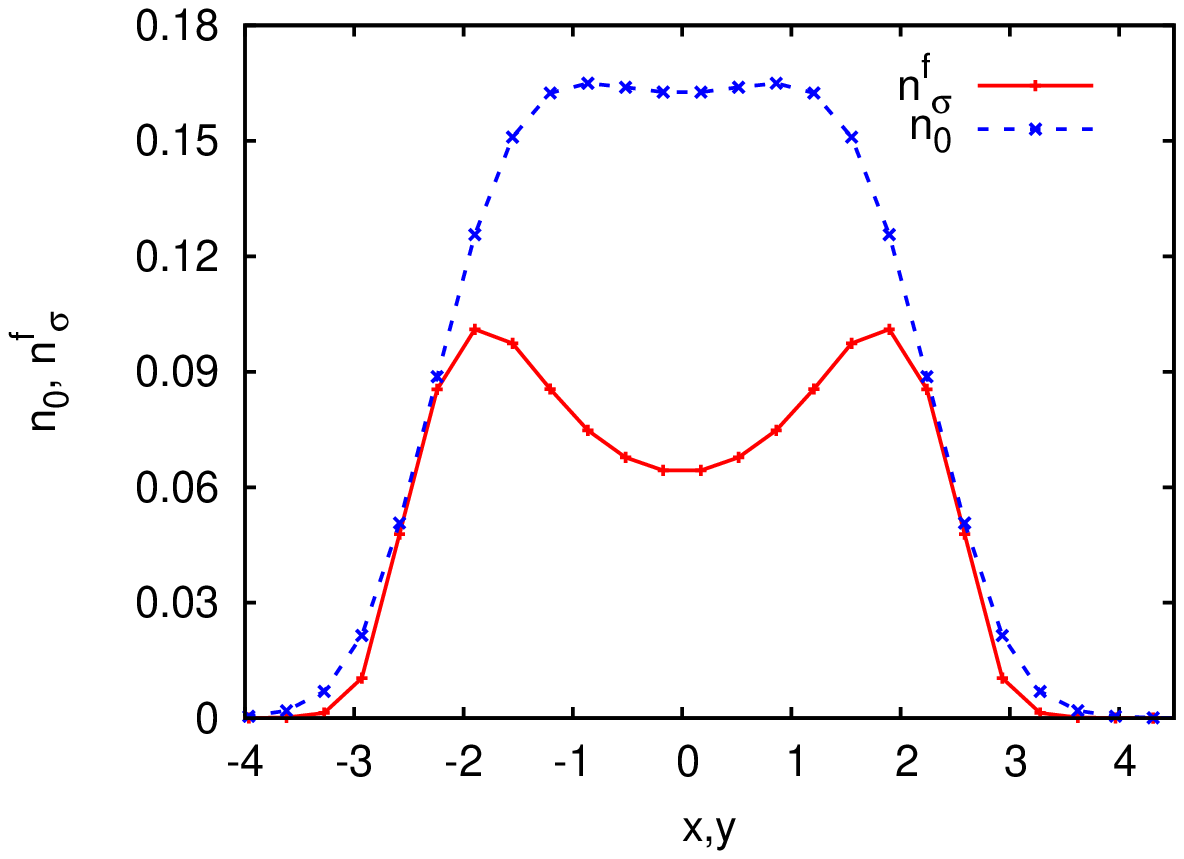}
\end{center}
\caption{The condensate density and the density of dissociated atoms
for the same values of $t/J$ and $U$ as in Fig. \ref{dens-U} but for higher densities with
$\mu_{\sigma}=0.1J$.
The dip deepens since at higher densities the role of the lattice is more profound.
The density profiles are along the lines $y=a/2$ and $x=a/2$.}
\label{fig:mu+0.1}
\end{figure}

Thus, the  contribution of the nonlocal nabla terms in Eqs. (\ref{E1}) and (\ref{E2}), which originate from the 
tunneling of the molecules in Eq. (\ref{model}), is essential for the appearance of the dip. 
The tunneling of such dressed molecules is due to an optical lattice and their fermionic nature \cite{duan05}.
So the appearance of the dip can be explained tentatively by the Pauli principle which acts between paired and 
unpaired fermions and causes the paired fermions to repel the 
dissociated ones (similar mechanism acts for unbalanced Fermi mixtures \cite{melo08}).
Eq. (\ref{model}) describes a one band fermionic
model and thus is valid for a deep optical periodic potential. For a shallow lattice
the Pauli principle is less efficient and we expect the dip formation to be less pronounced.
The Pauli principle is also less efficient if the density of fermions is small. We may expect that
in the dilute regime the role of the lattice is almost irrelevant.
Indeed, if we decrease the chemical potential, i.e., by going into the dilute
regime, the dip softens (cf. Fig. \ref{fig:mu-0.1}), while increasing the chemical potential makes the dip more
profound (cf. Fig. \ref{fig:mu+0.1}). Thus, the role of the lattice is crucial in the formation of the dip in the 
density of unpaired fermions at the trap center.





In conclusion, we have studied strongly attractive fermions in a 2D optical lattice superimposed by a trapping 
potential. We solved numerically the field equations obtaining non-homogeneous
densities of condensed molecules and dissociated fermions. The latter reveal an effect of dip
formation at the trap center. 
This can possibly be observed in future experiments and may serve as a signature of approaching of the strongly
coupling BEC regime of the BEC-BCS  crossover in a lattice.

O.F. and K.Z. acknowledge support from DAAD. A.G. thanks support from CAPES (Brazil) and FAPESP/CNPq (Brazil).

\appendix
\section{Summation over Matsubara frequencies}
\label{summat}
Here we show how we perform the summation over Matsubara frequencies. 
We transform $i\omega_n\rightarrow z$.
$G^{-1}(z)G^{-1}(z)$ is a Hermitian matrix ($i\Delta\equiv i\phi+\chi= i\bar{\phi}+\bar{\chi}$,
$\phi$ is real, $\chi$ is complex):
\begin{equation}
G^{-1}(z)G^{-1}(z)=\left(\begin{array}{cc} 
-\Delta^2+z^2-\mu^2-\hat{t}\mu-\mu\hat{t}-\hat{t}\hat{t} 
&
\hat{t}i\Delta-i\Delta\hat{t} \\ \hat{t}i\Delta-i\Delta\hat{t} &
-\Delta^2+z^2-\mu^2-\hat{t}\mu-\mu\hat{t}-\hat{t}\hat{t}
\end{array}\right).
\label{prodhermit}
\end{equation}
Any Hermitian matrix can be diagonalized by a unitary matrix:
\begin{equation}
G^{-1}(z)G^{-1}(z)=\hat{U}^{\dagger}\hat{\lambda}(z)\hat{U}, \ \ \hat{U}^{\dagger}=\hat{U}^{-1},
\label{iden}
\end{equation}
where $\hat{\lambda}$ is a diagonal (real) eigenvalue matrix. The equation for
$\hat{\lambda}$ of Eq.(\ref{prodhermit}) reads
\beq
\lambda_k=z^2-z_k^2.
\eeq
Reversing Eq. (\ref{iden}) we get
\beq
G(z)=\hat{U}^{\dagger}\frac{1}{z^2-z_k^2}\hat{U}G^{-1}(z).
\eeq
For the well behaved $g(x)$ (see \cite{simons06})
\beq
-\frac{1}{\beta}\sum_n \frac{g(i\omega_n)}{\omega_n^2+z_k^2}=\frac{1}{2\pi i}\oint dz \frac{g(z)f(z)}{z^2-z_k^2}=
\frac{g(z_k)f(z_k)-g(-z_k)f(-z_k)}{2z_k},
\eeq
where $f(z)=1/(e^{\beta z}+1)$. This works for the blocks $11$ and $22$ of the Green matrix in Eq. (\ref{Green}), 
since the function under the integral is well-behaved then. 
For the blocks $12$ and $21$ the sum is formally divergent since for large $\omega_n$ it behaves as $\sim 
1/\omega_n$. To cure the problem we must introduce convergent factors \cite{simons06,randeria}.

Using the above formula and the remark we get
\begin{equation}
\frac{1}{\beta}\sum_n G_{rr}(\omega_n)
=\sum_{km}\frac{\hat{U}_{rk}^{\dagger
}\hat{U}_{km}}{2z_k}\{\tanh[-\beta z_k/2]G_{mr}^{-1}(0)-z_k i \sigma_2\}.
\end{equation}


\end{document}